\documentclass[letterpaper]{article}

\usepackage{textcomp}
\usepackage{amsmath}
\usepackage{amssymb}
\usepackage{tabularx}
\usepackage{booktabs}
\usepackage[print-unity-mantissa=false, separate-uncertainty=true]{siunitx}
\usepackage{derivative}
\usepackage{bm}
\usepackage{mathtools}
\usepackage{isomath}
\usepackage{multirow}

\AddToHook{bfseries}{\boldmath}
\usepackage[T1]{fontenc}
\usepackage{geometry}
\usepackage{setspace}
\usepackage[style = chem-acs]{biblatex}
\usepackage{graphicx}
\usepackage{float}
\newfloat{scheme}{htbp}{los}
\floatname{scheme}{Scheme}
\floatname{chart}{Chart}
\newfloat{graph}{htbp}{loh}

\usepackage{chemformula} 
\usepackage[version = 4]{mhchem} 

\usepackage{authblk}
\author[1]{Zhijie Xu}
\author[2]{Ernesto Pini}
\author[2]{Lorenzo Pattelli*}
\affil[1]{Photonic Innovations Lab\\ Department of Electronic and Electrical Engineering\\ 
University College London\\
London WC1E 7JE, UK\\}
\affil[2]{Istituto Nazionale di Ricerca Metrologica\\ European Laboratory for Non-linear Spectroscopy \\
Sesto Fiorentino 50019, Italy\\}
\author[1]{Ioannis Papakonstantinou*}

\title{Revealing and Mitigating Transmittance and Reflectance Errors in Integrating-Sphere Characterization of High-Albedo Scattering Media}
\date{*Email: l.pattelli@inrim.it, i.papakonstantinou@ucl.ac.uk}
\input{macros.tex}

\begin{document}

\maketitle

\section*{Keywords}

Scattering, Point Spread Function, Monte Carlo Modelling, Optical Characterisation, Transmittance, Reflectance

\begin{abstract}
Accurate optical characterization of high-albedo scattering media is fundamentally limited by integrating-sphere measurement errors. Scattering events spatially redistribute photons, leading to lateral beam spread and angular divergence that bias measured reflectance and transmittance. Here, we investigate these deviations, which are quantified and interpreted by using the point spread function (PSF), guided by the radiative transfer equation. Experiments and Monte Carlo modelling show that the widely used external-mount integrating-sphere configuration introduces systematic, configuration-dependent errors that become pronounced when scattering-induced lateral spreading is substantial. Using a normalised PSF to quantify photon redistribution, we map the characteristic lateral spreading length across a broad $(\taus,g)$ parameter space in the high-albedo limit, providing a basis for evaluating errors in specific measurement geometries. We further quantify the influence of measurement geometry and show that, in the pencil-beam limit, maintaining $L/D_{\mathrm{port}}\lesssim 0.01$ reduces the discrepancy arising from the finite collection aperture to approximately \qty{1}{\percent}. In addition, we introduce an internal-mount integrating-sphere protocol that suppresses scattering-related losses and yields transmittance and reflectance measurements in good agreement with theoretical predictions. Overall, our work helps users of integrating-sphere measurements identify scattering-related errors and improve the reliability of transmittance and reflectance characterization across a wide range of application areas.
\end{abstract}

\section{Introduction}

Optical characterization based on spectral measurements of transmittance and reflectance is a foundational technique for evaluating material properties in a wide range of photonic applications. These include radiative cooling, luminescent solar concentrators (LSCs), solid-state lighting, display technologies, and biomedical imaging systems~\cite{raman2014passive, ramirez2024infrared, tummeltshammer2016losses, leyre2014absolute, wen2023ultrahigh, kim2013achieving, barbora2021higher}. In many of these fields, accurately detecting even minor amounts of absorption is paramount. Since absorption is typically inferred from transmittance and reflectance, any inaccuracies in primary measurements can propagate and compromise the accuracy of derived optical parameters. 

Light scattering, arising from embedded particles, surface roughness, or internal heterogeneity, introduces substantial uncertainty into transmittance and reflectance measurements~\cite{bohren2008absorption, berne2000dynamic}. Scattering events can broaden the transmitted and reflected intensity profiles beyond the physical dimensions of the detector's input aperture, causing part of the light to actually miss it. Meanwhile, scattering redistributes energy in a wide angular range; in strongly diffusive samples, the reflected photons can become highly randomized in direction, even when the incident beam is strongly collimated~\cite{mitchell2024energy, ishimaru1978wave}. As a result, a significant fraction of photons may lie outside the numerical aperture, or collection cone, of the detector. Together, these spatial and angular mismatches can lead to a substantial underestimation of transmittance and reflectance.

The use of an integrating sphere significantly mitigates measurement errors arising from angular divergence. Its inner surface is coated with a highly reflective diffuse layer that enables the collection of nearly the full solid angle, ensuring that photons transmitted and reflected at any angle ultimately contribute to the detected hemispherical signal~\cite{hanssen2001integrating}. However, while integrating sphere techniques alleviate angular mismatch, they still suffer from errors introduced by lateral beam spread. In scattering samples, lateral beam spread may cause transmitted or reflected light to miss the finite port aperture, resulting in underestimation, as shown in Figure~\ref{fig:first fig}. These effects coexist and compound with traditional integrating-sphere errors such as substitution errors, cavity recycling, and geometry-dependent effects~\cite{gevaux2020evaluating, foschum2020precise, bergmann2020precise, jonsson2003light, tao2024tutorial, frenzel2025influence}.

Recent efforts have improved the estimation of absorption and scattering coefficients through analytical, Monte Carlo, and inverse-modelling approaches~\cite{tao2024tutorial, foschum2020precise}, and have optimized integrating-sphere configurations to enhance measurement precision~\cite{bergmann2020precise}. Related studies have also modelled collection losses in finite-aperture reflectance/transmittance measurements of scattering materials, highlighting how beam spreading can lead to systematic underestimation~\cite{gevaux2020evaluating}.

However, beyond these conventional system-level errors, the extent to which a sample’s optical and geometrical parameters influence scattering events remains insufficiently quantified. A point spread function (PSF) describes this spatial redistribution and can be used to quantify the characteristic lateral spreading length, thereby linking port-related collection bias in reflectance and transmittance measurements~\cite{ishimaru1978wave}. In principle, the radiative transfer equation (RTE) provides an accurate description of the PSF. The scattering transport is characterized by the scattering optical thickness $\taus=\mus L$ and anisotropy $g$, where $\mus$ is the scattering coefficient and $L$ is the slab thickness \cite{bohren2008absorption, elaloufi2004diffusive, lemieux1998diffusing}. To bridge different scattering regimes comprehensively, Monte Carlo modelling provides a description of photon redistribution spanning from ballistic to diffusive transport without requiring an explicit solution of the RTE.

In this work, measurement errors in commonly used external-mount integrating-sphere systems are directly revealed through visual evidence of lateral beam spread and inconsistencies among experimental results. To predict the occurrence of these errors, we analyse the transmitted and reflected PSFs based on the RTE. The scattering transport, and consequently the PSF, is governed by the scattering coefficient $\mu_s$, absorption coefficient $\mu_a$, anisotropy factor $g$, and slab thickness $L$, for a given material system and boundary condition. Using the scaling invariance of the RTE, we employ a normalised PSF that depends on three dimensionless parameters: the scattering optical thickness $\taus$, the absorption optical thickness $\taua=\mua L$, and $g$. We focus on the high-albedo limit, which represents a worst-case scenario for scattering-related transmittance and reflectance errors because absorption suppresses long photon paths and thereby reduces lateral spreading. We therefore provide normalised PSF maps over a broad $(\taus,g)$ parameter space, quantified by the dimensionless radius $u_{95}$ enclosing \qty{95}{\percent} of the PSF energy, as a descriptor of the characteristic lateral spreading length. We further quantify how measurement bias in external-mount integrating-sphere systems depends on sample geometry and port aperture size $D_{\mathrm{port}}$. Under the investigated pencil-beam conditions, $L/D_{\mathrm{port}}\lesssim0.01$ provides a practical reference for limiting the relative finite-port error to approximately \qty{1}{\percent} across the investigated scattering conditions. In addition, we introduce a multi-step measurement procedure based on an internal-mount integrating-sphere configuration that suppresses scattering-related losses. For the \qty{2.5}{\milli\metre}-thick samples considered here, the proposed internal-mount protocol yields maximum absolute discrepancies of \qty{1.3}{\percent} in transmittance and \qty{6.0}{\percent} in reflectance relative to the corresponding simulations. For comparison, measurements of the same samples using the custom external-mount configuration show maximum absolute discrepancies of \qty{25.9}{\percent} and \qty{35.5}{\percent} in transmittance and reflectance, respectively. These guidelines are particularly relevant to commercial external-mount integrating-sphere systems, such as ultraviolet--visible--near-infrared (UV--Vis--NIR) spectrometers and Fourier-transform infrared (FTIR) systems, and provide practical guidance for minimising scattering-related errors and improving measurement accuracy using standard setups.

\section{Background and Motivation}

\subsection{Photon Pathways in a Scattering System}

For clarity, we define the entrance surface as the surface illuminated by the incident beam, as shown in Figure~\ref{fig:first fig} (a). The exit surface is the surface opposite to the entrance surface. Accordingly, photons leaving from the entrance surface contribute to the measured reflectance, whereas photons leaving from the exit surface contribute to the measured transmittance. We use the grey rectangle on the entrance surface to represent the reflection port of the integrating sphere, and the grey rectangle on the exit surface to represent the transmission port. Thus, the purple arrows passing within the grey regions are collected by the integrating sphere and contribute to the transmittance/reflectance measurements. However, when photons are scattered, not all of them fall within the port area, leading to measurement errors. Edge loss denotes photons that escape through the lateral surfaces of the sample, as indicated by the dark-blue arrows in Figure~\ref{fig:first fig} (a). In addition, some photons leaving the sample from the entrance/exit surfaces (which would otherwise contribute to reflectance/transmittance) miss the ports, as shown by the yellow and red arrows in Figure~\ref{fig:first fig} (a). We refer to these losses as port losses, and distinguish R-port loss and T-port loss for the reflection and transmission sides, respectively.

\subsection{Sample-Size Induced Errors}

\begin{figure}
  \includegraphics[width=\linewidth]{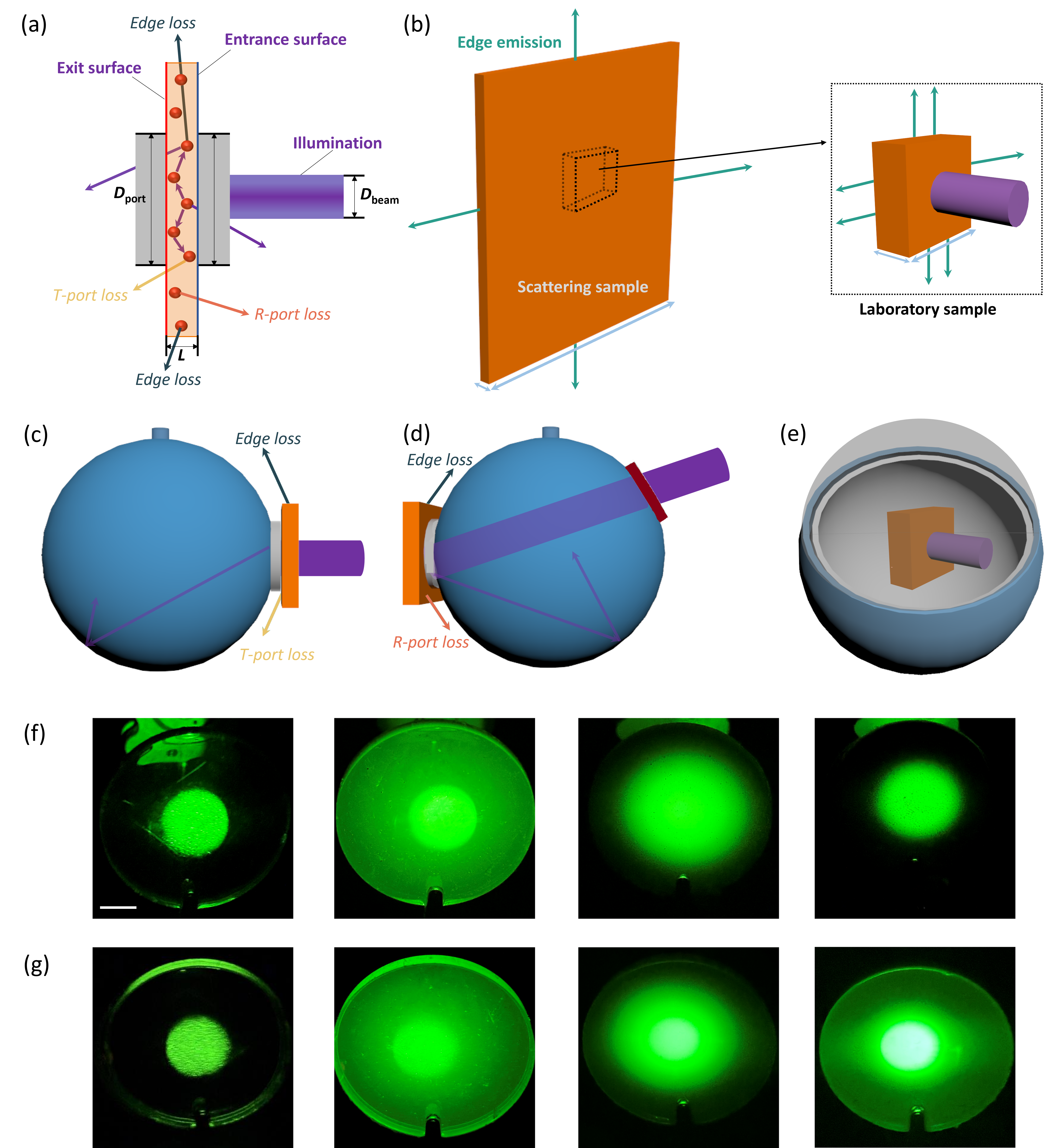}
  \caption{(a) Schematic of photon pathways and measurement losses. Grey rectangles denote integrating-sphere ports and red spheres denote scatterers. Red/blue lines mark the exit/entrance surfaces. Purple arrows indicate collected photons; dark-blue arrows indicate edge loss; yellow/red arrows indicate transmission/reflection port loss. $L$, $D_{\text{port}}$, and $D_{\text{beam}}$ denote the sample thickness, port aperture diameter, and incident beam diameter, respectively.
(b) Edge loss in large-scale versus laboratory-scale samples. 
(c) Transmittance and (d) reflectance measurements using an external-mount integrating sphere, with the errors indicated by teal, red and yellow arrows.
(e) Configuration of an internal-mount integrating sphere.
(f) Transmitted and (g) reflected intensity profiles versus PND (left to right: \qtylist[print-unity-mantissa=true,list-units=single]
{1e6;1e7;1e8;1e9}
{\per\milli\meter\cubed}). With increasing PND, both profiles first broaden and then gradually narrow, illustrating the non-monotonic dependence of lateral beam spread on scattering conditions.}
  \label{fig:first fig}
\end{figure}

In the development of materials, laboratory measurements are often performed on downscaled prototypes that retain the same thickness as real-world samples but have substantially smaller lateral dimensions, while still being large enough to fully cover the integrating-sphere port aperture, as illustrated in Figure~\ref{fig:first fig} (b). The laboratory sample can be regarded as a small piece cut from the much larger final product, as indicated by the dashed frame. For example, radiative cooling materials are typically intended to cover large roofs with areas on the order of several square meters, whereas samples tested in laboratory-scale studies usually measure only a few square centimetres \cite{raman2014passive, li2025accelerated}. This compact format is commonly used to reduce material consumption and testing time during early-stage evaluation. Besides, in practice, it is unrealistic to test meter-scale samples using laboratory instruments such as UV--Vis--NIR because of the size limitations of the detection setup. 

However, in scattering media, such lateral downscaling can introduce a representativeness error arising from the finite sample size itself. For samples with the same scattering parameters, including $\mus$, $\mua$, and $g$, and the same thickness $L$, the intrinsic photon redistribution remains the same, as discussed in detail in the Point Spread Function in Scattering Media Section. However, reducing the lateral dimensions of the sample introduces additional lateral boundaries through which redistributed photons can escape. Consequently, photons that escape through the edges of a finite laboratory sample would remain within a laterally extended material, where they may continue to propagate and eventually escape through either the entrance or exit surface. The transmittance and reflectance of the finite sample can therefore differ from those of the large-area material that it is intended to represent, introducing a finite-size representativeness error.
 
In addition to geometrical constraints, the refractive index contrast between the scattering medium and its surroundings also plays an important role. Fresnel reflection and light guiding within the sample, resulting from total internal reflection (TIR), scattering-induced transport or anisotropic material morphology, can further enhance edge and port losses. The influence of surface roughness may also affect lateral photon redistribution, although its quantitative modelling and impact still represent an open question that goes beyond the scope of this paper.

\subsection{System-Induced Errors}
\label{system error}

In addition to sample-size-induced effects, measurement errors can arise from limitations of the measurement setup itself. Integrating spheres are widely used to measure total reflectance and transmittance from scattering samples because they collect light over nearly the entire solid angle~\cite{hanssen2001integrating}. However, their finite port apertures and sample configuration can still introduce port and edge losses. 

Current systems are generally classified into two geometries based on sample positioning: external mount (Figure~\ref{fig:first fig} (c) and (d)) and internal mount (Figure~\ref{fig:first fig} (e) also referred to as ``centre mount'' \cite{storm1998use}). In this paper, we use these terms as a naming convention for clarity, although they are not standardized in the literature. Most commercial spectroscopic instruments utilize the external-mount configuration for compactness and ease of alignment. In commercial systems, key parameters such as port aperture, sample positioning, and incident beam size are usually fixed by the manufacturer. Figures~\ref{fig:first fig} (c) and (d) illustrate the working principles and associated limitations of external-mount setups for reflectance and transmittance, respectively. To visually illustrate lateral beam spread and potential leakage from the opening of the external-mount integrating sphere, we captured images of four samples with different concentration of scatterers. As shown in Figures~\ref{fig:first fig} (f) and (g), the transmitted and reflected intensity profiles exhibit pronounced lateral redistribution, with their spatial extent varying non-monotonically across the investigated scattering conditions. In particular, substantial emission from the sample edges is visible at a particle number density (PND) of \qty[print-unity-mantissa=true]{1e7}{\per\milli\meter\cubed} for both transmission and reflection, directly illustrating the lateral escape of redistributed photons that gives rise to edge loss in the measurements. Lateral redistribution can additionally lead to finite-port loss when photons emerging from the entrance or exit surface fall outside the collection aperture, as quantitatively investigated in the following sections. For clarity, since a reflectance setup typically has two ports, the discussion of aperture size for both transmittance and reflectance in this paper refers only to the port connected to the sample, as indicated by the grey rectangles in Figures~\ref{fig:first fig} (c) and (d).

Regarding the internal-mount integrating sphere, it is mostly employed to characterize the overall properties of a sample, such as photoluminescence quantum yield (PLQY) or luminescence lifetime using time-correlated single-photon counting (TCSPC). In this configuration, because the sample is mounted inside the sphere, it is difficult to distinguish photons originating from different surfaces, which makes it challenging to separate transmittance, reflectance, and edge loss. Nevertheless, the internal-mount configuration also offers a clear advantage: placing the sample inside the sphere substantially mitigates systematic errors related to limited port apertures. Building on this advantage, we propose a method to measure transmittance and reflectance with an internal-mount integrating sphere, which will be discussed in the Internal Mount Integrating Sphere Section.

\subsection{Point Spread Function in Scattering Media}
\label{sec: PSF}

Photon redistribution in a scattering system can be described by the PSF, which can be obtained from the RTE for the specific intensity. We consider steady-state transport in a homogeneous slab occupying $0\le z\le L$ with infinite lateral extent. With no internal sources present, the specific intensity $I(\mathbf{r},\boldsymbol{\Omega})$ (\unit{\watt\per\meter\squared\per\steradian}), with $\mathbf{r}=(\boldsymbol{\rho},z)$ where $\boldsymbol{\rho}$ is the transverse position vector and $z$ is the axial coordinate, and $\boldsymbol{\Omega}$ the unit propagation-direction vector ($|\boldsymbol{\Omega}|=1$), satisfies \cite{lenoble1985radiative, ishimaru1978wave, chandrasekhar2013radiative}
\begin{equation}
\boldsymbol{\Omega}\cdot\nabla I(\mathbf{r},\boldsymbol{\Omega})
+(\mua + \mus)\,I(\mathbf{r},\boldsymbol{\Omega})
=
\mus\int_{4\pi} p(\boldsymbol{\Omega}'\!\cdot\boldsymbol{\Omega};g)\,
I(\mathbf{r},\boldsymbol{\Omega}') \odif{\boldsymbol{\Omega}'},
\label{eq:RTE_dim}
\end{equation}
where $p(\cdot;g)$ (\unit{\per\steradian}) is a normalized phase function with $\int_{4\pi}p \odif{\boldsymbol{\Omega}}=1$ \cite{van2012multiple}, and $g=\langle\cos\Theta\rangle\in[-1,1]$, with $\Theta$ as the scattering angle between the incoming and outgoing propagation directions \cite{jacques2013optical}. 

We use a normally incident pencil beam at $z=0^+$ \cite{patterson1989time}, with injected radiant power $P_0$ (\unit{\watt}),
\begin{equation}
I(\boldsymbol{\rho},0,\boldsymbol{\Omega})\big|_{\boldsymbol{\Omega}\cdot\hat{\mathbf z}>0}
=
P_0\,\delta^{(2)}(\boldsymbol{\rho})\,\delta (\boldsymbol{\Omega}-\hat{\mathbf z}),
\label{eq:BC_pencil}
\end{equation}
where $\delta^{(2)}(\boldsymbol{\rho})$ (\unit{\per\square\meter}) is the 2D Dirac delta in the transverse plane, enforcing a point-like input at $\boldsymbol{\rho}=\mathbf{0}$, and $\delta(\boldsymbol{\Omega}-\hat{\mathbf z})$ (\unit{\per\steradian}) is the angular Dirac delta enforcing normal, collimated incidence.

We define the PSF on $z=z_{\text{surf}}$ as the outward hemispherically integrated normal flux density (\unit{\watt\per\meter\squared}) \cite{tuchin2015tissue},
\begin{equation}
\mathrm{PSF}(\boldsymbol{\rho})
=
\int_{\mathcal H}
\big|\boldsymbol{\Omega}\cdot\hat{\mathbf z}\big|\,
I(\boldsymbol{\rho},z_{\text{surf}},\boldsymbol{\Omega})
\odif{\boldsymbol{\Omega}},
\label{eq:PSF_def}
\end{equation}
where $z_{\text{surf}}=0$ for reflection and $z_{\text{surf}}=L$ for transmission, and the outward hemisphere is
$\mathcal H_-=\{\boldsymbol{\Omega}:\boldsymbol{\Omega}\cdot\hat{\mathbf z}<0\}$ or
$\mathcal H_+=\{\boldsymbol{\Omega}:\boldsymbol{\Omega}\cdot\hat{\mathbf z}>0\}$, respectively.

The intrinsic dimensionless parameters are \cite{ishimaru1978wave}
\begin{equation}
\taus=\mus L,\qquad \taua=\mua L,\qquad g.
\label{eq:taus}
\end{equation}
Using the incident power $P_0$ (\unit{\watt}) to normalize the amplitude, the PSF admits the scaling form \cite{bouchaud1990anomalous}
\begin{equation}
\mathrm{PSF}(\rho)
=
\frac{P_0}{L^2}\,
f\!\left(\frac{\rho}{L};\taus,\taua,g\right),
\label{eq:PSF_scaling}
\end{equation}
where $f$ is the normalised PSF. Equation~\eqref{eq:PSF_scaling} makes the parameter dependence explicit: for any given refractive index of the material under study, the PSF shape is governed solely by $(\taus,\taua,g)$, while changing $L$ at fixed $(\taus,\taua,g)$ produces a self-similar lateral stretching by $L$ and an amplitude reduction proportional to $1/L^2$ (thickness-scaling effect). With $f$ defined in Eq.~\eqref{eq:PSF_scaling}, the exit-plane irradiance (outward normal flux density) for a finite-diameter beam follows by 2D convolution, \cite{tuchin2015tissue}
\begin{equation}
E_{\text{out}}(\boldsymbol{\rho})
=
\frac{P_0}{L^2}\,
\Big(E_{\text{in}} * f\!\big(|\boldsymbol{\rho}|/L;\taus,\taua,g\big)\Big)(\boldsymbol{\rho}),
\label{eq:conv_f}
\end{equation}
where $*$ denotes 2D convolution in the transverse plane.

\subsection{Point Spread Function across Varying $\taus$ and $g$}

We conducted a series of Monte Carlo simulations (Monte Carlo tool MCX \cite{fang2022mcx}),
spanning multiple parameter combinations by using a pencil beam. In all simulations, the scatterers were monodisperse particles with a fixed radius embedded in a polydimethylsiloxane (PDMS) transparent host matrix. Unless otherwise stated, the optical parameters $\taus$, $\taua$, and $g$ used in the following simulations and comparisons are evaluated at \qty{550}{\nano\metre}. This wavelength was selected because absorption by the TiO$_2$--PDMS system is negligible, allowing the high-albedo limit considered here to be examined experimentally. Exploiting the thickness-scaling effect, lateral beam spread can be fully characterized by the normalised PSF, which depends only on $(\taus,\taua,g)$. In the high-albedo regime ($\gtrsim 0.99$), which is both common in practice and typically of greatest technological relevance \cite{jacques2013optical}, the dependence on absorption becomes negligible and the normalised PSF can be further simplified as:

\begin{equation}
\mathrm{PSF}(\rho)
=
\frac{P_0}{L^2}\,
f\!\left(\frac{\rho}{L};\taus,g\right),
\qquad \taua \approx 0.
\label{eq:psf_high_albedo}
\end{equation}

\begin{figure}
  \centering \includegraphics[width=\linewidth]{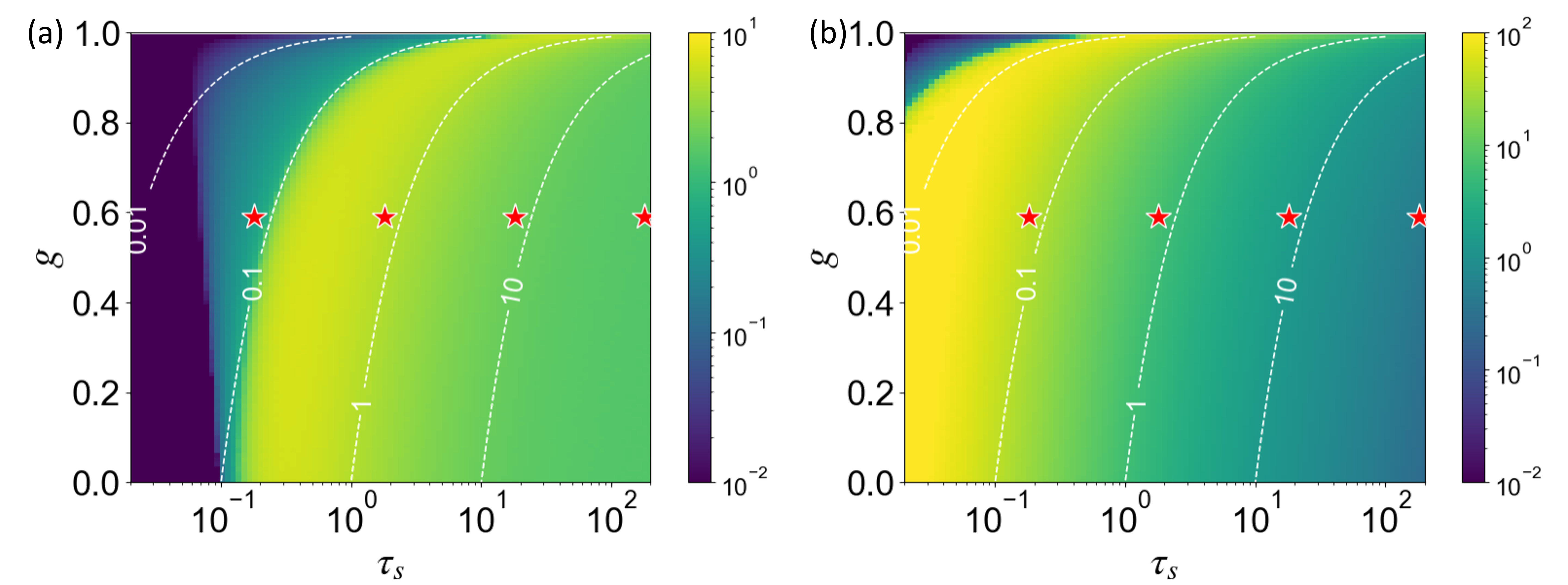}
  \caption{Parameter sweep of lateral beam spread quantified by the dimensionless radius $u_{95}$ as a function of $\taus$ and $g$: (a) transmission and (b) reflection. White dashed lines indicate contours of constant reduced scattering optical thickness, $\tausp=\taus(1-g)$. The star symbols denote the four conditions selected for the subsequent simulations and experiments at $g=0.59$, with $\taus=\numlist[list-units=single]{0.18;1.8;18;180}$.}
  \label{fig: sweep}
\end{figure}

To span a broad range of $\taus$ and $g$, we designed a set of simulation conditions with $\taus$ varying from \num{0.02} to \num{200} and $g$ from \num{0.01} to \num{0.99},  thereby covering scattering regimes ranging from ballistic to diffusive transport \cite{ishimaru1978wave, patterson1989time, van2012multiple, jacques2013optical}. The outcomes are summarized using $u_{95}=\rho_{95}/L$, where $\rho_{95}$ is the radius enclosing \qty{95}{\percent} of the transmitted or reflected power. Figure~\ref{fig: sweep} (a) and (b) show $u_{95}$ for transmission and reflection, respectively.

Figure~\ref{fig: sweep} (a) and (b) reveal a pronounced, non-monotonic dependence of $u_{95}$ on $\taus$. For transmission, $u_{95}$ increases with $\taus$ and reaches its largest values at approximately $0.3\lesssim\taus\lesssim3$, indicating maximal lateral spreading when scattering is sufficient to substantially redistribute photons but not yet strong enough to restrict their long-distance propagation. As $\taus$ increases beyond this range, $u_{95}$ decreases and approaches a smaller, slowly varying plateau. In contrast, for reflection, the strongest lateral broadening generally occurs at smaller $\taus$ values, with the precise location depending on $g$, and $u_{95}$ decreases markedly as $\taus$ increases. Overall, these heatmaps provide a practical lookup map of the characteristic lateral spreading length across the investigated $(\taus,g)$ parameter space, with clearly different ``worst-case'' conditions for transmitted and reflected PSFs. This non-monotonic evolution is qualitatively consistent with the transmitted and reflected intensity profiles shown in Figure~\ref{fig:first fig} (f) and (g), which similarly broaden and subsequently narrow with increasing PND.

Because the reduced scattering optical thickness, $\tausp=\musp L$, provides a useful measure of the overall scattering regime, contours of constant $\tausp$ (\numlist{0.01;0.1;1;10}) are superimposed on Figure~\ref{fig: sweep} (a) and (b). Overall, the variation of $u_{95}$ broadly follows these contours, indicating that $\tausp$ captures the dominant scaling of the lateral spreading. The correspondence is not exact across the entire parameter space, with the clearest deviations occurring in the weak-scattering region of the transmission map, indicating a residual dependence on $\taus$ and $g$ separately. For $\tausp \gtrsim 1$, the colour distribution becomes more closely aligned with these contours, suggesting that $\tausp$ provides an increasingly effective description of the lateral spreading in the strongly multiple-scattering regime. This is because successive scattering progressively erases the directional memory of individual scattering events, making photon transport less sensitive to the specific combination of $\taus$ and $g$. 

Having established the general dependence of lateral spreading on $(\taus,g)$, we subsequently consider a fixed-$g$ trajectory through this parameter space. Figure~\ref{fig: sweep} shows that the influence of $g$ on $u_{95}$ is relatively modest over much of the investigated parameter space and becomes pronounced mainly toward the strongly forward-scattering limit. Moreover, $\taus$ can be systematically tuned through the scatterer concentration, whereas $g$ is primarily determined by the single-particle scattering properties of a given material system and is therefore less readily adjusted independently in practice. A fixed-$g$ trajectory thus provides a physically relevant and experimentally controllable route for examining the evolution of lateral spreading with $\taus$.

Specifically, we consider spherical TiO\textsubscript{2} nanoparticles with a radius of \qty{100}{\nano\metre} embedded in PDMS ($n=1.43$) at \qty{550}{\nano\metre}, for which absorption is negligible and the single-particle anisotropy factor is $g\approx0.59$. Varying the nanoparticle concentration therefore enables $\taus$ to be systematically varied along this fixed-$g$ trajectory, which spans a broad range of $u_{95}$ values in Figure~\ref{fig: sweep}.

The physical thickness was fixed at $L=\qty{2.5}{\milli\metre}$ unless otherwise stated, while the TiO\textsubscript{2} nanoparticle concentration was varied to obtain PNDs of \qtylist[print-unity-mantissa=true,list-units=single]
{1e6;1e7;1e8;1e9}
{\per\milli\meter\cubed}. These correspond to particle volume fractions (PVFs) of approximately \numlist{4.2e-6;4.2e-5;4.2e-4;4.2e-3}. Thus,  scattering optical thicknesses can be directly estimated as $\taus\approx\numlist{0.18;1.8;18;180}$, respectively, spanning three orders of magnitude in $\taus$. The corresponding $u_{95}$ values are marked in Figure~\ref{fig: sweep}. The scattering and absorption cross sections, together with the anisotropy factor, were calculated using Mie theory based on the optical constants of TiO\textsubscript{2} and PDMS. The results are provided in Figure~S1, and the complete set of optical parameters is summarized in Table~S1 of the Supporting Information. Representative PSFs for these conditions are shown in Figure~S2 to illustrate the corresponding variation in lateral beam spread.

\section{Results and Discussion}

\subsection{Edge-Loss Dependence on Sample Lateral Dimension}
\label{Edge Loss}
\begin{figure}
  \includegraphics[width=\linewidth]{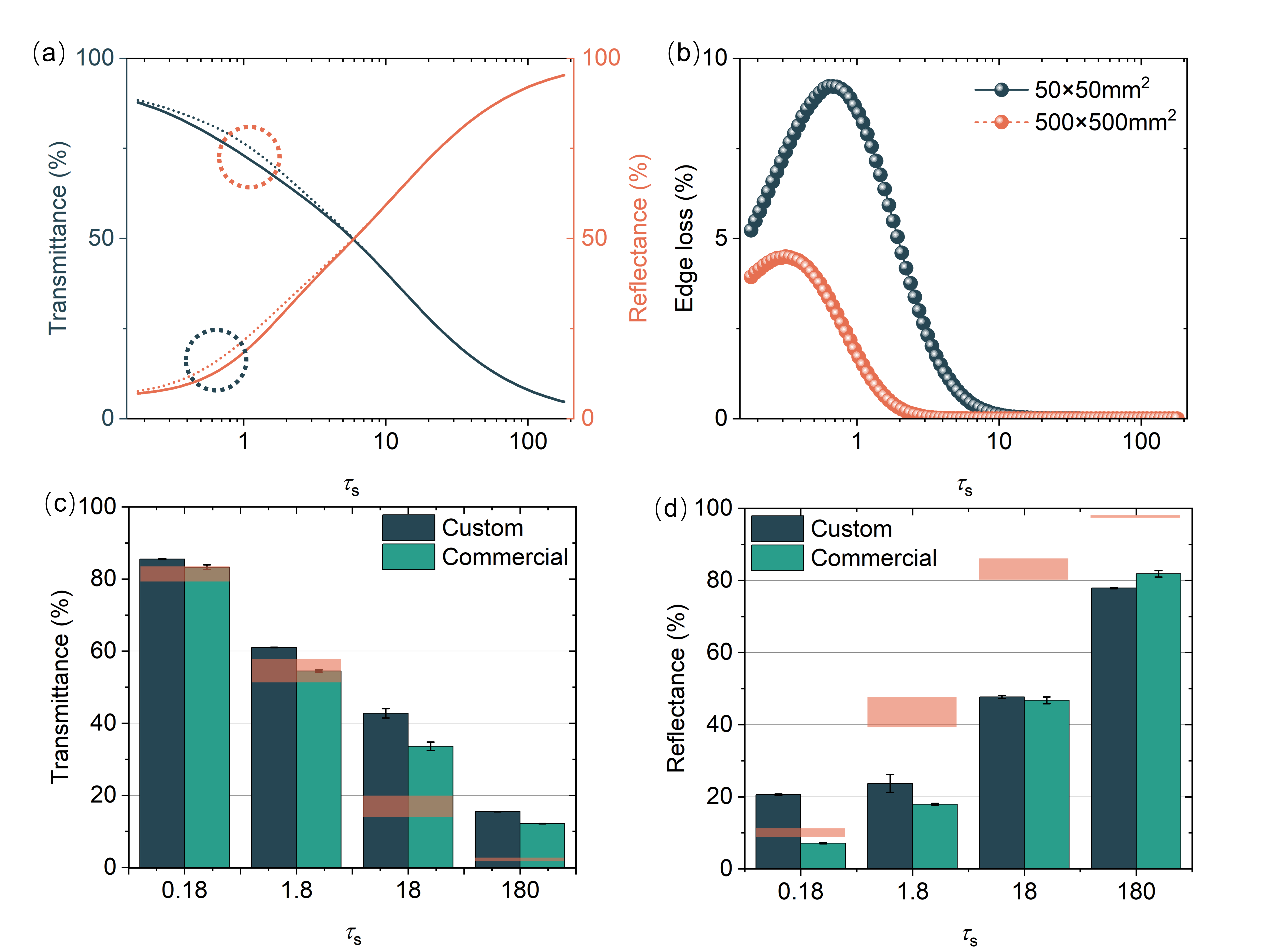}
  \caption{(a) Comparison of transmittance (dark blue lines) and reflectance (red lines) between a large-scale device (\qtyproduct[product-units=power]{500 x 500}{\milli\meter}, dotted lines) and a small tested sample (\qtyproduct[product-units=power]{50 x 50}{\milli\meter}, solid lines). The most apparent differences are highlighted with dashed circles.
  (b) Comparison of edge loss between the large-scale and laboratory-scale samples.
  (c) Transmittance and (d) reflectance of samples with different $\taus$, measured using a custom-built external-mount integrating sphere (dark blue) and a commercial external-mount integrating sphere (teal). Red squares represent the range of variation of simulated transmittance and reflectance.
  }
  \label{fig:edge loss}
\end{figure}

As discussed previously, discrepancies between large-scale and laboratory-scale samples can arise from differences in their lateral dimensions, primarily through edge loss. To quantify this effect, we simulated reflectance and transmittance for two sample sizes---\qtyproduct[product-units=power]{500 x 500}{\milli\meter} and \qtyproduct[product-units=power]{50 x 50}{\milli\meter}---across four representative $\taus$ values (Figure~\ref{fig:edge loss} (a)).

The simulation results reveal systematic differences in both reflectance and transmittance between the small and large samples, with the magnitude of these differences varying non-monotonically with $\taus$. For both quantities, the discrepancy initially increases and subsequently decreases as $\taus$ increases. The largest absolute transmittance difference, \qty{3.35}{\percent}, occurs at $\taus\approx\num{1}$, while the maximum absolute reflectance difference of \qty{3.23}{\percent} occurs at a slightly smaller value of $\taus\approx\num{0.65}$. At small $\taus$ ($\taus\lesssim\num{0.3}$), most photons undergo few or no scattering events, resulting in limited lateral beam spread and therefore relatively small edge loss. As $\taus$ increases to approximately \numrange[range-phrase=\text{--}]{0.3}{3}, lateral beam spread becomes more pronounced, allowing a larger fraction of photons to escape through the sample edges and increasing the difference between the two sample sizes. At larger $\taus$ ($\taus\gtrsim\num{3}$), the characteristic lateral spreading decreases, as shown in Figure~\ref{fig: sweep}, and the resulting edge loss is correspondingly reduced. Similar enhancement of in-plane photon diffusion has been observed in optically thin slabs, where lateral transport becomes increasingly significant relative to the slab thickness, thereby amplifying edge losses in small samples \cite{pini2024non}.

To isolate the role of edge loss, we simulated the total irradiance emitted from the edges of finite parallelepiped samples for both sample sizes at each $\taus$ (Figure~\ref{fig:edge loss} (b)). The edge loss exhibits a similar non-monotonic dependence on $\taus$, with the difference between the small and large samples becoming most pronounced at intermediate $\taus$ values, approximately between \num{0.3} and \num{3}. Over this range, the smaller sample exhibits substantially greater edge loss, consistent with the larger differences in measured reflectance and transmittance. The edge loss of the small sample reaches its maximum at $\taus\approx\num{1}$, whereas that of the large sample peaks at a smaller value of $\taus\approx\num{0.3}$. This shift of the edge-loss peak toward smaller $\taus$ for the larger sample can be understood from the competition between the characteristic lateral spreading length and the lateral sample dimension. As the sample lateral dimension increases, a larger characteristic lateral spreading length is required to achieve significant edge loss. This condition is met at a smaller $\taus$.
The quantitative edge losses reported here correspond to the refractive-index contrast used in our simulations ($n\approx1.43$ for the polymer host relative to air). Different host refractive indices will modify Fresnel reflection and total internal reflection at the sample boundaries and may therefore shift the magnitude of the edge loss, although the underlying finite-size mechanism remains unchanged \cite{pattelli2016diffusive, mazzamuto2016deducing}.

\subsection{Measured Transmittance and Reflectance Errors}
\label{Errors in IS}

To assess systematic errors in external-mount integrating-sphere measurements, we compared samples with varying $\taus$ using a custom-built integrating-sphere system (RTC-060-SF, Labsphere) and a commercial UV--Vis--NIR spectrophotometer equipped with an integrating-sphere attachment (UV-3600i Plus, Shimadzu), with Monte Carlo simulations providing the intrinsic transmittance and reflectance as theoretical references. All measurements were performed at \qty{550}{\nano\metre} using samples with a thickness of $L=\qty{2.5}{\milli\metre}$. For the custom-built system, the incident beam diameter was $D_{\text{beam}}=\qty{1}{\milli\metre}$ and the sample-facing port diameter was $D_{\text{port}}=\qty{5}{\milli\metre}$ for both transmittance and reflectance measurements, with reflectance measured at an incidence angle of approximately \qty{8}{\degree}. The corresponding beam and port dimensions of the commercial system are fixed by the instrument configuration. Details of the two measurement configurations are shown in Figure~S3, with the complete instrumentation and measurement procedures provided in the Experimental Section.

The simulated intrinsic values are not treated as strict ground truth because of uncertainties associated with both sample fabrication and modelling. Achieving homogeneous scatterer dispersions at the designed PNDs is experimentally challenging, particularly at low PNDs, where the required particle quantity is small and difficult to control accurately. At high PNDs, particle aggregation may also occur, whereas the Monte Carlo model assumes the nominal optical properties of isolated particles. To account for possible variations in the actual PND, additional simulations were performed with the PND varied by $\pm\qty{10}{\percent}$ around each target value, and the resulting range is reported as the theoretical reference. Further details of the simulations are provided in the Effect of Port Aperture on Measurement Accuracy Section.

The results, shown in Figure~\ref{fig:edge loss} (c) and (d), demonstrate that the magnitude of the discrepancies between the external-mount measurements and simulations strongly depends on $\taus$. For transmittance (Figure~\ref{fig:edge loss} (c)), substantial deviations are observed, with maximum absolute discrepancies of \qty{25.9}{\percent} and \qty{16.8}{\percent} for the custom-built and commercial external-mount systems, respectively, at $\taus=\num{18}$. Similarly, for reflectance (Figure~\ref{fig:edge loss} (d)), the maximum absolute discrepancies occur at $\taus=\num{18}$, reaching \qty{35.5}{\percent} and \qty{36.4}{\percent} for the custom-built and commercial external-mount systems, respectively. Further details regarding the measurements are provided in Figure~S4.

\subsection{Effect of Port Aperture on Measurement Accuracy}
\label{Effect of Port Aperture}

The discrepancies confirmed in the Measured Transmittance and Reflectance Errors Section are significantly affected by the port aperture. To quantify the resulting finite-port errors, we simulated and evaluated the fraction of power loss via Monte Carlo simulations. A disk-shaped source was used as the incident illumination. As the reference, total transmittance ($T_\text{true}$) and reflectance ($R_\text{true}$) were obtained using idealized detectors that are geometrically coincident with the sample surfaces and collect all power emitted from those surfaces, without any spatial or angular restriction.

To emulate realistic detection constraints, circular detectors with diameters of \qtylist[list-units=single]{2;5;10;20}{\milli\metre} were positioned on the entrance and exit surfaces of the sample. These apertures capture only a fraction of the emitted power, yielding the detected transmittance $T_\text{det}$ and reflectance $R_\text{det}$. The relative finite-port errors were then defined as:

\begin{equation}
\varepsilon_T = \frac{T_\text{true} - T_\text{det}}{T_\text{true}},
\label{rel_T}
\end{equation}

\begin{equation}
\varepsilon_R = \frac{R_\text{true} - R_\text{det}}{R_\text{true}},
\label{rel_R}
\end{equation}

\begin{figure}
  \includegraphics[width=\linewidth]{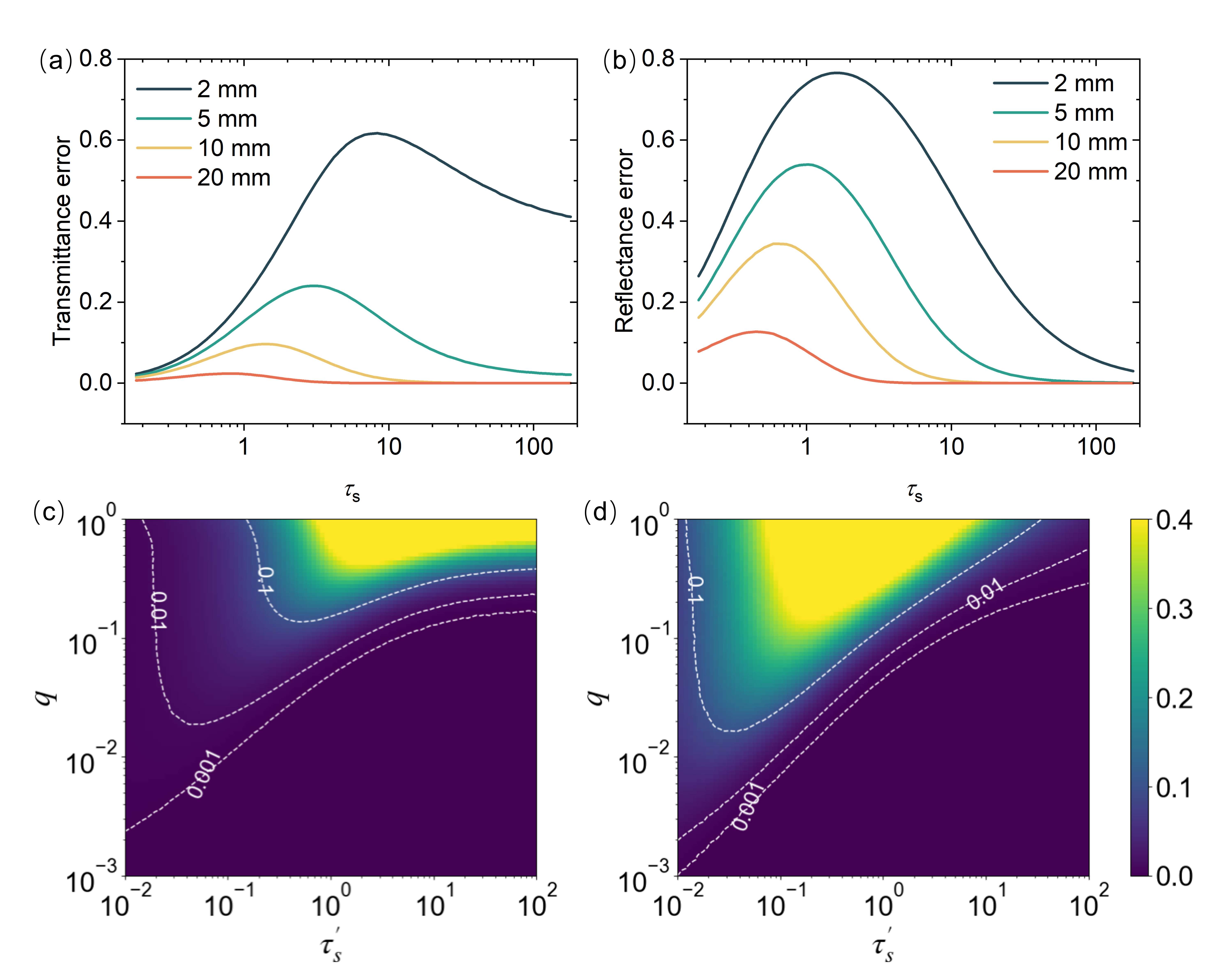}
  \caption{Relative finite-port errors in (a) transmittance and (b) reflectance for various virtual detector aperture diameters, representing the integrating-sphere port size at the sample-mounted port, using an incident beam diameter of \qty{2}{\milli\metre}.
  Relative finite-port errors in (c) transmission and (d) reflection as functions of $\tausp$ and $q=L/D_{\text{port}}$ under pencil-beam illumination, with $g=0.59$ and a slab refractive index of $n=1.43$. The colour scale indicates the relative finite-port error, while the white dashed contours correspond to error levels of \num{0.001}, \num{0.01}, and \num{0.1}.}
  \label{fig:spot-size}
\end{figure}

A schematic of the simulation geometry is provided in Figure~S5. All simulations assumed a sample thickness of \qty{2.5}{\milli\metre} and a central wavelength of \qty{550}{\nano\metre}. The incident intensity-profile diameter was set to \qty{2}{\milli\metre}, matching the smallest detector aperture. The spot-size variation of \qty{2}{\milli\metre} is simulated in Figure~S6 to capture the lateral beam spread. The dependence of the relative finite-port errors on scattering conditions is shown in Figure~\ref{fig:spot-size} (a) and (b). 

For transmittance, we observed that when the port aperture reached \qty{20}{\milli\metre}, approximately ten times the incident beam diameter, $\varepsilon_T$ remained below \qty{2.37}{\percent} across all $\taus$. In contrast, smaller apertures exhibited substantial losses around $\taus\approx\num{10}$, where $\varepsilon_T$ exceeded \qty{60}{\percent} for a port aperture of \qty{2}{\milli\meter}. This trend matches the transmitted beam evolution shown in Figure~S6 and the characteristic lateral spreading lengths in Figure~\ref{fig: sweep}: both indicate that transmitted intensity profiles undergo their strongest lateral expansion in this $\taus$ range (up to nearly five times the incident diameter), which maximizes the port loss. At larger $\taus$ ($\taus\gtrsim\num{30}$), transmitted intensity profiles contract again, reducing the loss and lowering the error.

For reflectance, $\varepsilon_R$ instead peaks at smaller $\taus$ values, approximately $\taus\approx\numrange[range-phrase=\text{--}]{0.3}{2}$. This behaviour is consistent with Figure~S6, where reflected intensity profiles show their most pronounced expansion at approximately $\taus\approx\numrange[range-phrase=\text{--}]{0.3}{3}$, with spot sizes exceeding ten times the incident diameter. The corresponding characteristic lateral spreading lengths in Figure~\ref{fig: sweep} are also largest in this range, and the strong angular divergence makes reflected photons particularly prone to escaping small apertures. As $\taus$ increases further, reflected intensity profiles shrink rapidly and their PSFs narrow, causing $\varepsilon_R$ to decrease. Across the investigated $\taus$ range, transmittance errors are smaller than reflectance errors, consistent with the much stronger reflected-lateral beam spread seen in Figure~S6. To directly visualize how lateral beam spread evolves with $\taus$ and how this spread impacts the accuracy of transmittance and reflectance measurements, representative photographs are further illustrated in Figure~S7. Additional irradiance simulations, provided in the Supporting Information (Figure~S8 and S9), further confirm the beam-size evolution described above.

To examine how finite-port errors scale with sample thickness and port size beyond the specific geometries considered above, we further consider the pencil-beam limit. The relative finite-port errors defined in Equations~\ref{rel_T} and~\ref{rel_R} are evaluated as functions of $\tausp$ and the dimensionless sample-thickness-to-port-diameter ratio
\[
q=\frac{L}{D_{\text{port}}}.
\]
These simulations were performed using a pencil-beam source, with a fixed anisotropy factor of $g=0.59$ and a slab refractive index of $n=1.43$. Under these conditions, the incident beam introduces no additional lateral length scale, allowing the geometrical dependence to be characterized by $q$.

Figure~\ref{fig:spot-size} (c) and (d) show the relative finite-port errors in transmission and reflection, respectively. In both cases, the error remains small at sufficiently low $q$ and increases as $q$ becomes larger. The largest errors occur at intermediate $\tausp$, where lateral beam spread is most pronounced, while the errors decrease towards both the low- and high-$\tausp$ limits. The white dashed contours indicate relative finite-port errors of \num{0.001}, \num{0.01}, and \num{0.1}. Under the investigated pencil-beam conditions, $q\lesssim\num{0.01}$ limits the relative finite-port error to approximately \qty{1}{\percent} across the investigated $\tausp$ range for both transmission and reflection. The corresponding absolute finite-port errors are provided in the Supporting Information, where $q\lesssim\num{0.02}$ limits the absolute error to approximately \qty{1}{\percent} across the investigated range.

The $q$ dependence in Figure~\ref{fig:spot-size} (c) and (d) should therefore be interpreted as a pencil-beam reference rather than a universal criterion for finite-width illumination. For a finite incident beam, the output distribution additionally depends on the incident beam profile through its convolution with the pencil-beam PSF, and the beam size relative to the port aperture must therefore also be considered.

\subsection{Internal Mount Integrating Sphere}
\label{Internal Mount Integrating Sphere}
\begin{figure}
  \includegraphics[width=\linewidth]{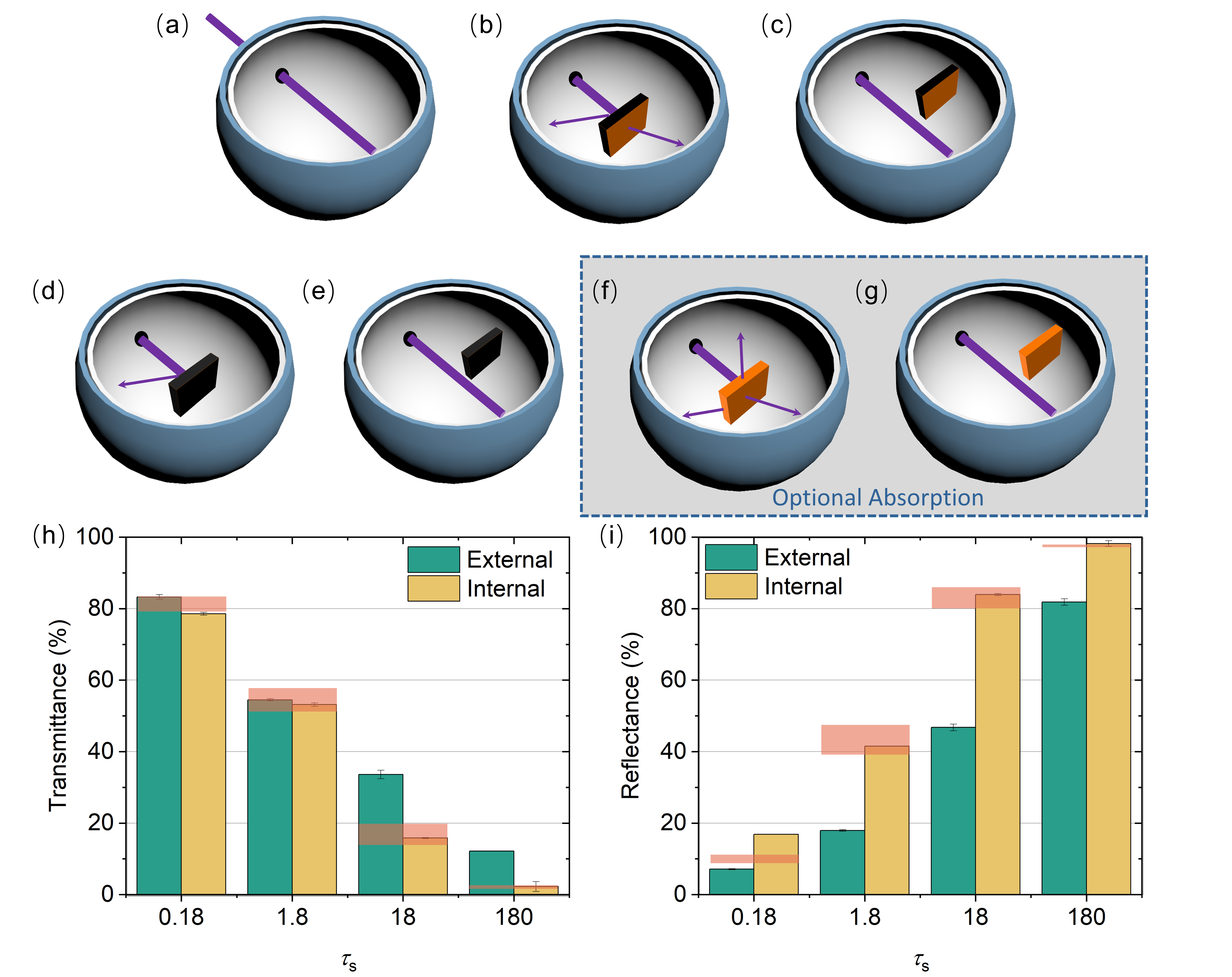}
  \caption{Schematic of the internal-mount integrating sphere protocol. 
(a) Reference measurement with the empty sphere ($I_\text{empty}$). 
(b) Sample with side edges blackened and directly illuminated ($I_\text{edge,dir}$). 
(c) Same as (b), but shielded from direct illumination to record the cavity-induced background ($I_\text{edge,ind}$). 
(d) Sample with both side edges and exit surface blackened and directly illuminated ($I_\text{back,dir}$). 
(e) Same as (d), but shielded from direct illumination to record the cavity-induced background ($I_\text{back,ind}$). 
(f) Sample inserted and directly illuminated ($I_\text{sample,dir}$). 
(g) Sample inserted, but shielded from direct illumination to record the cavity-induced background ($I_\text{sample,ind}$). 
(f) and (g) are included to enable potential absorption measurements. 
(h) Comparison of measured transmittance and (i) reflectance as a function of $\taus$ using external- and internal-mount configurations. The red squares represent the range of variation of the simulations for reference.}
  \label{fig:internal}
\end{figure}

Based on the above discussion, lateral beam spread and angular divergence can lead to both port and edge losses in external-mount integrating-sphere measurements of scattering samples. These losses can be substantially reduced by placing the sample inside the integrating sphere, where photons emerging from different surfaces of the sample can be collected more completely without the constraint imposed by a finite sample-facing port. The remaining challenge is then to distinguish photons emerging from the different sample surfaces. To achieve this, we employ an internal-mount integrating-sphere apparatus together with a multi-step measurement procedure. A related internal-mount apparatus has previously been used to quantify edge losses in LSCs \cite{tummeltshammer2016losses}, but without separately resolving transmittance and reflectance. Here, we extend this approach by introducing measurement configurations that separately determine transmittance, reflectance, and edge loss, as shown in Figure~\ref{fig:internal} (a--g).

In the first step (Figure~\ref{fig:internal} (a)), a reference measurement is obtained with the empty sphere ($I_\text{empty}$).

In the second step (Figure~\ref{fig:internal} (b--c)), the sample is inserted with its side edges blackened by a high-absorption coating (optical absorption \qty{>99}{\percent}) to suppress lateral leakage. Paired measurements are then performed with and without direct illumination, giving
\begin{equation}
I_\text{edge}^{\text{corr}} = I_\text{edge,dir} + \bigl(I_\text{empty} - I_\text{edge,ind}\bigr).
\end{equation}
The indirect-illumination measurement is introduced to reduce the interference caused by the sample itself, sometimes referred to as substitution error in standard internal-mount configurations \cite{tummeltshammer2016losses,bergmann2020precise,foschum2020precise}.

In the third step (Figure~\ref{fig:internal} (d--e)), the exit surface of the sample is additionally blackened to suppress back-reflected contributions. Again, paired measurements are performed with and without direct illumination, yielding
\begin{equation}
I_\text{back}^{\text{corr}} = I_\text{back,dir} + \bigl(I_\text{empty} - I_\text{back,ind}\bigr).
\end{equation}

For the high-albedo samples considered in this work, absorption is negligible to a good approximation. Under this condition, the simplified protocol in Figure~\ref{fig:internal} (a--e) is sufficient to estimate the main measured quantities. The transmittance is obtained as
\begin{equation}
\text{Transmittance} = \frac{I_\text{edge}^{\text{corr}} - I_\text{back}^{\text{corr}}}{I_\text{empty}},
\end{equation}
while the edge loss is approximated as
\begin{equation}
\text{Edge Loss} \approx \frac{I_\text{empty} - I_\text{edge}^{\text{corr}}}{I_\text{empty}}.
\end{equation}
The reflectance is then estimated as
\begin{equation}
\text{Reflectance} \approx 1 - \text{Transmittance} - \text{Edge Loss}.
\end{equation}

If the sample has non-negligible absorption, two additional measurements can be included, as shown in Figure~\ref{fig:internal} (f--g). In this case, the sample is inserted without blackening the exit surface, and paired measurements with and without direct illumination give
\begin{equation}
I_\text{sample}^{\text{corr}} = I_\text{sample,dir} + \bigl(I_\text{empty} - I_\text{sample,ind}\bigr).
\end{equation}
The absorption can then be estimated as
\begin{equation}
\text{Absorption} = \frac{I_\text{empty} - I_\text{sample}^{\text{corr}}}{I_\text{empty}}.
\end{equation}

Using this protocol, we re-measured the samples and compared the results with those obtained from commercial external-mount configurations, as shown in Figure~\ref{fig:internal} (h) and (i), where the yellow bars represent the results obtained using the proposed protocol and the teal bars represent the commercial external-mount measurements. Overall, both transmittance and reflectance achieved using this protocol show good agreement with the simulations, as indicated by the red squares in Figure~\ref{fig:internal} (h) and (i). With the exception of two reflectance measurements, all measured values fall within the corresponding simulation variation ranges. At $\taus\gtrsim30$, the internal-mount reflectance is \qty{98.16}{\percent}, only slightly above the simulated range of \qtyrange{97.34}{98.06}{\percent}. A larger discrepancy occurs at the smallest $\taus$ considered, where the measured reflectance is \qty{16.85}{\percent}, compared with the simulated range of \qtyrange{8.47}{10.86}{\percent}. This mismatch most likely arises from overcorrection of the integrating-sphere response using the indirect-illumination measurements in Figure~\ref{fig:internal} (c) and (e). At small $\taus$, the incident beam undergoes relatively few scattering events and is therefore only weakly redistributed by the sample, so the indirect-illumination correction does not simply remove a small sample-induced perturbation, but can instead overcompensate for it. A similar effect was also reported in low-concentration LSC measurements \cite{tummeltshammer2016losses}. By contrast, in conventional external-mount integrating-sphere measurements, absolute discrepancies can exceed \qty{20}{\percent} for transmittance and \qty{35}{\percent} for reflectance at larger $\taus$.

\subsection{Practical Guidelines for External-Mount Integrating-Sphere Measurements}

When an internal-mount configuration is unavailable, scattering-related errors in external-mount integrating-sphere measurements can be reduced by jointly considering the sample thickness, lateral dimensions, incident beam size, and port aperture.

The sample thickness should be kept as small as permitted by the measurement objective. Because the dimensional lateral beam spread scales with $L$ for fixed dimensionless optical parameters, reducing $L$ generally restricts the distance over which photons can spread laterally. However, changing $L$ also changes $\taus=\mus L$ for a given material; reducing the sample thickness should therefore be regarded as a practical geometrical measure rather than a universal correction.

The lateral sample dimensions should be sufficiently large relative to the illuminated region and the resulting photon redistribution to minimize edge loss, with the incident beam positioned near the sample centre. The port aperture should likewise be sufficiently large to collect the broadened output distribution. The pencil-beam simulations in Figure~\ref{fig:spot-size} (c) and (d) provide a limiting reference for this requirement, with $q=L/D_{\text{port}}\lesssim0.01$ limiting the relative finite-port error to approximately \qty{1}{\percent} under the investigated conditions. For finite-width illumination, however, the incident beam introduces an additional lateral length scale. As illustrated in Figure~S11, convolution of finite-width input beams with the corresponding pencil-beam PSFs produces progressively broader output distributions as the incident beam diameter increases. The relative increase in output size is most pronounced for narrow incident beams, where scattering-induced spreading dominates, whereas for broader beams the output size becomes increasingly determined by the incident beam itself. Thus, the pencil-beam criterion should be regarded as a limiting reference rather than a universal design rule: in practical measurements, the incident beam profile and the scattering-induced PSF jointly determine the output footprint that must be accommodated by both the sample and the collection aperture.

Accordingly, reducing the incident beam size can provide additional spatial margin for scattering-induced spreading, but beam size should not be considered independently of the sample and port dimensions. The sample should remain sufficiently larger than the broadened output footprint to suppress edge loss, while the port should provide sufficient collection margin to suppress finite-port loss. Increasing the port aperture is nevertheless constrained by the allowable port fraction of the integrating sphere and by other losses associated with an excessively large sphere opening. The practical objective is therefore to balance sample size, beam size, and port aperture so that the expected output distribution is adequately contained without unnecessarily increasing the sphere opening.

For broadband measurements, $\taus$, $\taua$, and $g$ vary with wavelength; therefore, the PSF and the resulting scattering-related measurement errors can also vary spectrally. These geometrical considerations should consequently be evaluated over the relevant wavelength range rather than at a single wavelength when broadband accuracy is required.

\section{Conclusion}

In summary, we investigate scattering-related errors in integrating-sphere measurements by linking spatial photon redistribution to transmittance and reflectance biases. Analysis of the RTE shows that lateral beam spread follows a self-similar scaling law with sample thickness $L$ and can be described by a normalised PSF. Combined Monte Carlo modelling and experiments demonstrate that scattering-induced lateral redistribution can lead to substantial port and edge losses in conventional external-mount integrating-sphere configurations. A $(\taus,g)$-resolved PSF map quantifies the characteristic lateral spreading length across a broad parameter space and provides a practical basis for identifying conditions susceptible to measurement bias. We further quantify the influence of measurement geometry and show that, under the investigated pencil-beam conditions, $L/D_{\mathrm{port}}\lesssim 0.01$ provides a practical reference for limiting relative finite-port errors to approximately \qty{1}{\percent}. To more effectively suppress scattering-related losses, we introduce an internal-mount integrating-sphere protocol. For the high-albedo samples considered here, this protocol yields maximum absolute discrepancies of \qty{1.3}{\percent} in transmittance and \qty{6.0}{\percent} in reflectance relative to the corresponding simulations, compared with \qty{25.9}{\percent} and \qty{35.5}{\percent}, respectively, using the custom external-mount configuration. Together, these results provide both practical guidance for reducing errors in conventional external-mount measurements and an internal-mount alternative when higher measurement accuracy is required.


\section{Experimental Section}
\label{Experimental section}
\threesubsection{Preparation of Scattering Samples}

To achieve better dispersion of TiO\textsubscript{2} nanoparticles with a nominal radius of \qty{100}{\nano\meter} and to ensure compatibility with the PDMS matrix, the nanoparticles were first dissolved in a mixed solvent composed of isopropyl alcohol (IPA) and hexane in a 1:1 volume ratio. The resulting TiO\textsubscript{2} suspension was ultrasonicated for \qty{30}{\minute} to obtain a well-dispersed stock solution. Based on the desired PND, appropriate volumes of the TiO\textsubscript{2} stock solution were extracted and mixed with the PDMS elastomer. The mixture was stirred using a magnetic stirrer to ensure uniformity and subsequently subjected to ultrasonic treatment for one hour at a controlled temperature of \qty{6}{\celsius}. After sonication, a curing agent (Sylgard 184, Dow Corning) was added to the mixture at a weight ratio of 1:10 (curing agent:elastomer). The final mixture was degassed and poured into pre-fabricated plastic molds. The molds were placed on a hot plate at \qty{60}{\celsius} for \qty{3}{\hour}. After curing, the samples were demolded to yield the final scattering samples.

\threesubsection{External-Mount Integrating Sphere}

A custom-built external-mount integrating-sphere system (RTC-060-SF, Labsphere) was used to emulate the working principle of commercial external-mount integrating-sphere systems for measuring transmittance and reflectance. Illumination was provided by a 6335 light source (Newport) and spectrally selected at \qty{550}{\nano\metre} using a monochromator (Oriel Instruments). The incident beam diameter was $D_{\text{beam}}=\qty{1}{\milli\metre}$, and the sample-facing port diameter was $D_{\text{port}}=\qty{5}{\milli\metre}$ for both transmittance and reflectance measurements. Light collected by the integrating sphere was coupled to a spectrometer (Ocean Optics) using a \qty{2}{\metre}-long step-index multimode fibre patch cable (M71L02, Thorlabs) with SMA905 connectors, a \qty{1000}{\micro\metre} core diameter, and a numerical aperture of 0.48. Unlike commercial instruments, the custom-built system allowed the beam and port dimensions to be varied independently.

To measure transmittance, the incident beam was first directed into the integrating sphere without a sample, and the corresponding signal $I_{\text{beam}}$ was recorded. The sample was then placed at the front opening of the integrating sphere, and the transmitted signal $I_{\text{trans}}$ was measured inside the sphere. The transmittance $T$ was calculated as
\[
T = \frac{I_{\text{trans}}}{I_{\text{beam}}}.
\]

For reflectance measurements, a nominal \qty{99}{\percent} Spectralon diffuse reflectance standard (USRS-99-010, AS-01158-060, Labsphere) was first placed at the sample port and illuminated at an incidence angle of approximately \qty{8}{\degree}. The reference signal $I_{\text{standard}}$ was recorded. The standard was then replaced with the sample under the same illumination geometry, and the reflected signal $I_{\text{ref}}$ was collected. The sample reflectance $R$ was calculated as
\[
R = \frac{I_{\text{ref}}}{I_{\text{standard}}} R_{\text{standard}},
\]
where $R_{\text{standard}}$ is the reflectance of the reference standard. All measurements were corrected by subtracting the background signal obtained from a dark measurement acquired using the same integration time. 

\threesubsection{Internal-Mount Integrating-Sphere Measurements}

The internal-mount measurements were performed using the integrating-sphere apparatus reported previously in Ref.~\cite{tummeltshammer2016losses}. The apparatus consisted of a \qty{50}{\centi\metre}-diameter integrating sphere (Mulansphere) illuminated by a \qty{100}{\watt} quartz tungsten halogen light source (6335, Newport) coupled to a monochromator. Light from the monochromator was coupled into an optical fibre and collimated into the integrating sphere using a collimating lens (Ocean Optics). Light collected by the sphere was coupled through an optical fibre to a CCD spectrometer (Ocean Optics). All measurements were performed at \qty{550}{\nano\metre}.

For measurement configurations requiring selective suppression of photons leaving particular sample surfaces, the corresponding surfaces were coated with a highly absorbing black coating (Musou Black, water-based acrylic paint). All measurements were corrected by subtracting the background signal obtained from a dark measurement acquired using the same integration time.

\section*{Acknowledgements}

I.P. and Z.X. thank ERC-PoC-PolyCool and UKRI Grant No. EP/X024482/1, ERC-PoC-CoolGlass and UKRI Grant No. EP/Y036603/1, HorizonEurope SINCERE and InnovateUK Grant No. 10092236, and Pump-Priming funding and UCL Grand Challenge of Climate Crisis on Food Security Grant No.\ 589256 for funding. L.P. and E.P. acknowledge support from the project 21GRD03 PaRaMetriC. The project 21GRD03 PaRaMetriC has received funding from the European Partnership on Metrology, co-financed from the European Union’s Horizon Europe Research and Innovation Programme and by the Participating States. L.P. acknowledges support from the project FREECOOL, which received funding from the Ministero dell’Universit\`{a} e della Ricerca (MUR) under the frame of the Horizon Europe Cofund LEAP-SE, (GA N° 101172838).

\section*{Supporting information}

Additional simulation details, experimental methods, and supplementary figures are provided in the Supporting Information.

\printbibliography

\end{document}